\input epsf
%%%%%%%%%%%%%%%%%%%%%%%%%%%%%%%%%%%%%%%%%%%%%%%%%%%%%%%%%%%%%%%%%
%                                                               %
%       FONT FAMILIES:                                          %
%                                                               %
%%%%%%%%%%%%%%%%%%%%%%%%%%%%%%%%%%%%%%%%%%%%%%%%%%%%%%%%%%%%%%%%%
%                                                                %
%       Define script letters as rsfs                           %
%               (or redefine as cal)                            %
%                                                               %
%                                                               %
%%%%%%%%%%%%%%%%%%%%%%%%%%%%%%%%%%%%%%%%%%%%%%%%%%%%%%%%%%%%%%%%%
\newfam\scrfam
\batchmode\font\tenscr=rsfs10 \errorstopmode
\ifx\tenscr\nullfont
        \message{rsfs script font not available. Replacing with calligraphic.}
        \def\scr{\cal}
\else   
        \font\sevenscr=rsfs7
        \font\fivescr=rsfs5
        \skewchar\tenscr='177 \skewchar\sevenscr='177 \skewchar\fivescr='177
        \textfont\scrfam=\tenscr \scriptfont\scrfam=\sevenscr
        \scriptscriptfont\scrfam=\fivescr
        \def\scr{\fam\scrfam}
        \def\cal{\scr}
\fi
%%%%%%%%%%%%%%%%%%%%%%%%%%%%%%%%%%%%%%%%%%%%%%%%%%%%%%%%%%%%%%%%%
%                                                               %
%       fraktur (or redefine as italic)		                %
%                                                               %
%%%%%%%%%%%%%%%%%%%%%%%%%%%%%%%%%%%%%%%%%%%%%%%%%%%%%%%%%%%%%%%%%
\catcode`\@=11
\newfam\frakfam
\batchmode\font\tenfrak=eufm10 \errorstopmode
\ifx\tenfrak\nullfont
        \message{eufm font not available. Replacing with italic.}
        
\else
	
	\font\sevenfrak=eufm7 \font\fivefrak=eufm5
	\textfont\frakfam=\tenfrak
	\scriptfont\frakfam=\sevenfrak \scriptscriptfont\frakfam=\fivefrak
	
\fi
\catcode`\@=\active
%%%%%%%%%%%%%%%%%%%%%%%%%%%%%%%%%%%%%%%%%%%%%%%%%%%%%%%%%%%%%%%%%
%                                                               %
%       Blackboard bold (or redefine as boldface)               %
%                                                               %
%%%%%%%%%%%%%%%%%%%%%%%%%%%%%%%%%%%%%%%%%%%%%%%%%%%%%%%%%%%%%%%%%
\newfam\msbfam
\batchmode\font\twelvemsb=msbm10 scaled\magstep1 \errorstopmode
\ifx\twelvemsb\nullfont\def\Bbb{\bf}

	\message{Blackboard bold not available. Replacing with boldface.}
\else   \catcode`\@=11
        \font\tenmsb=msbm10 \font\sevenmsb=msbm7 \font\fivemsb=msbm5
        \textfont\msbfam=\tenmsb
        \scriptfont\msbfam=\sevenmsb \scriptscriptfont\msbfam=\fivemsb
        \def\Bbb{\relax\expandafter\Bbb@}
        \def\Bbb@#1{{\Bbb@@{#1}}}
        \def\Bbb@@#1{\fam\msbfam\relax#1}
        \catcode`\@=\active

\fi
%%%%%%%%%%%%%%%%%%%%%%%%%%%%%%%%%%%%%%%%%%%%%%%%%%%%%%%%%%%%%%%%%
%                                                               %
%       More FONTS:                                             %
%                                                               %
%%%%%%%%%%%%%%%%%%%%%%%%%%%%%%%%%%%%%%%%%%%%%%%%%%%%%%%%%%%%%%%%%
        \font\eightrm=cmr8              \def\xrm{\eightrm}
        \font\eightbf=cmbx8             \def\xbf{\eightbf}
        \font\eightit=cmti10 at 8pt     \def\xit{\eightit}
%%%     \font\eightit=cmti8             \def\xit{\eightit}
        \font\eighttt=cmtt8             
        
        \font\eighti=cmmi8              \def\xold{\eighti}
        \font\eightib=cmmib8             \def\xbold{\eightib}
        \font\teni=cmmi10               \def\old{\teni}

        \font\twelvecp=cmcsc10 scaled\magstep1

	 at10pt	
	\font\twelvehelvbold=phvb at12pt
	 at14pt
	\font\sixteenhelvbold=phvb at16pt

	\font\twelvebf=cmb12
	\font\twelveit=cmti12

\def\noblackbox{\overfullrule=0pt}
\noblackbox

%%%%%%%%%%%%%%%%%%%%%%%%%%%%%%%%%%%%%%%%%%%%%%%%%%%%%%%%%%%%%%%%%
%                                                               %
%       HEADLINE:                                               %
%                                                               %
%%%%%%%%%%%%%%%%%%%%%%%%%%%%%%%%%%%%%%%%%%%%%%%%%%%%%%%%%%%%%%%%%
\newtoks\oddheadtext
\newtoks\evenheadtext
\headline={\ifnum\pageno=1\hfill\else
	\ifodd\pageno{\eightrm\the\oddheadtext}{ }\dotfill{ }{\old\folio}
	\else{\old\folio}{ }\dotfill{ }{\eightrm\the\evenheadtext}\fi
	\fi}
\def\makeheadline{\vbox to 0pt{\vss\noindent\the\headline\break
\hbox to\hsize{\hfill}}
        \vskip2\baselineskip}
%%%%%%%%%%%%%%%%%%%%%%%%%%%%%%%%%%%%%%%%%%%%%%%%%%%%%%%%%%%%%%%%%
%                                                               %
%       FOOTNOTES:                                              %
%                                                               %
%%%%%%%%%%%%%%%%%%%%%%%%%%%%%%%%%%%%%%%%%%%%%%%%%%%%%%%%%%%%%%%%%
\newcount\infootnote
\infootnote=0
\def\foot#1#2{\infootnote=1
\footnote{${}^{#1}$}{\vtop{\baselineskip=.75\baselineskip
\advance\hsize by -\parindent\noindent{\xrm #2}\hfill\vskip2pt}}\infootnote=0$\,$}
%%%%%%%%%%%%%%%%%%%%%%%%%%%%%%%%%%%%%%%%%%%%%%%%%%%%%%%%%%%%%%%%%
%                                                               %
%       REFERENCES:                                             %
%                                                               %
%%%%%%%%%%%%%%%%%%%%%%%%%%%%%%%%%%%%%%%%%%%%%%%%%%%%%%%%%%%%%%%%%
\newcount\refcount
\refcount=1
\newwrite\refwrite
\def\oldsize{\ifnum\infootnote=1\xold\else\old\fi}
\def\ref#1#2{
	\def#1{{{\oldsize\the\refcount}}\ifnum\the\refcount=1\immediate\openout\refwrite=\jobname.refs\fi\immediate\write\refwrite{\item{[{\xold\the\refcount}]} 
	#2\hfill\par\vskip-2pt}\xdef#1{{\noexpand\oldsize\the\refcount}}\global\advance\refcount by 1}
	}
\def\refout{\catcode`\@=11
        \xrm\immediate\closeout\refwrite
        \vskip2\baselineskip
        {\noindent\twelvecp References}\hfill\vskip\baselineskip
                                                %\vskip.25\baselineskip%%%%
        %\parskip=.875\parskip
        %\baselineskip=.8\baselineskip
        \baselineskip=.75\baselineskip
        \input\jobname.refs
        %\parskip=8\parskip \divide\parskip by 7
        %\baselineskip=1.25\baselineskip
        \baselineskip=4\baselineskip \divide\baselineskip by 3
        \catcode`\@=\active\rm}

\def\hepth#1{\href{http://xxx.lanl.gov/abs/hep-th/#1}{hep-th/{\xold#1}}}
\def\jhep#1#2#3#4{\href{http://jhep.sissa.it/stdsearch?paper=#2\%28#3\%29#4}{J. High Energy Phys. {\xbold #1#2} ({\xold#3}) {\xold#4}}}
\def\AP#1#2#3{Ann. Phys. {\xbold#1} ({\xold#2}) {\xold#3}}

\def\CMP#1#2#3{Commun. Math. Phys. {\xbold#1} ({\xold#2}) {\xold#3}}
\def\CQG#1#2#3{Class. Quantum Grav. {\xbold#1} ({\xold#2}) {\xold#3}}
\def\HPA#1#2#3{Helv. Phys. Acta {\xbold#1} ({\xold#2}) {\xold#3}}

\def\JHEP{\jhep}

\def\MPLA#1#2#3{Mod. Phys. Lett. {\xbf A}{\xbold#1} ({\xold#2}) {\xold#3}}
\def\NPB#1#2#3{Nucl. Phys. {\xbf B}{\xbold#1} ({\xold#2}) {\xold#3}}

\def\PLB#1#2#3{Phys. Lett. {\xbf B}{\xbold#1} ({\xold#2}) {\xold#3}}

\def\PRD#1#2#3{Phys. Rev. {\xbf D}{\xbold#1} ({\xold#2}) {\xold#3}}

%%%%%%%%%%%%%%%%%%%%%%%%%%%%%%%%%%%%%%%%%%%%%%%%%%%%%%%%%%%%%%%%%
%                                                               %
%       SECTION NUMBERING:                                      %
%                                                               %
%%%%%%%%%%%%%%%%%%%%%%%%%%%%%%%%%%%%%%%%%%%%%%%%%%%%%%%%%%%%%%%%%
\newcount\sectioncount
\sectioncount=0
\def\section#1#2{\global\eqcount=0
	\global\subsectioncount=0
        \global\advance\sectioncount by 1
	\ifnum\sectioncount>1
	        \vskip\baselineskip
	\fi
	\noindent
        \line{\twelvebf\the\sectioncount. #2\hfill}
		\vskip.2\baselineskip\noindent
        \xdef#1{{\old\the\sectioncount}}}
\newcount\subsectioncount
\def\subsection#1#2{\global\advance\subsectioncount by 1
	\vskip.8\baselineskip\noindent
	\line{\tenbf\the\sectioncount.\the\subsectioncount. #2\hfill}
	\vskip.5\baselineskip\noindent
	\xdef#1{{\old\the\sectioncount}.{\old\the\subsectioncount}}}
\newcount\appendixcount
\appendixcount=0
\def\appendix#1{\global\eqcount=0
        \global\advance\appendixcount by 1
        \vskip2\baselineskip\noindent
        \ifnum\the\appendixcount=1
        \hbox{\twelvecp Appendix A: #1\hfill}\vskip\baselineskip\noindent\fi
    \ifnum\the\appendixcount=2
        \hbox{\twelvecp Appendix B: #1\hfill}\vskip\baselineskip\noindent\fi
    \ifnum\the\appendixcount=3
        \hbox{\twelvecp Appendix C: #1\hfill}\vskip\baselineskip\noindent\fi}
\def\acknowledgements{\vskip2\baselineskip\noindent
        \underbar{\it Acknowledgements:}\ }
%%%%%%%%%%%%%%%%%%%%%%%%%%%%%%%%%%%%%%%%%%%%%%%%%%%%%%%%%%%%%%%%%
%                                                               %
%       EQUATION NUMBERING                                      %
%                                                               %
%%%%%%%%%%%%%%%%%%%%%%%%%%%%%%%%%%%%%%%%%%%%%%%%%%%%%%%%%%%%%%%%%
\newcount\eqcount
\eqcount=0
\def\Eqn#1{\global\advance\eqcount by 1
\ifnum\the\sectioncount=0
	\xdef#1{{\noexpand\oldsize\the\eqcount}}
	\eqno({\oldstyle\the\eqcount})
\else
        \ifnum\the\appendixcount=0
	        \xdef#1{{\noexpand\oldsize\the\sectioncount}.{\noexpand\oldsize\the\eqcount}}
                \eqno({\oldstyle\the\sectioncount}.{\oldstyle\the\eqcount})\fi
        \ifnum\the\appendixcount=1
	        \xdef#1{{\oldstyle A}.{\old\the\eqcount}}
                \eqno({\oldstyle A}.{\oldstyle\the\eqcount})\fi
        \ifnum\the\appendixcount=2
	        \xdef#1{{\oldstyle B}.{\old\the\eqcount}}
                \eqno({\oldstyle B}.{\oldstyle\the\eqcount})\fi
        \ifnum\the\appendixcount=3
	        \xdef#1{{\oldstyle C}.{\old\the\eqcount}}
                \eqno({\oldstyle C}.{\oldstyle\the\eqcount})\fi
\fi}
\def\eqn{\global\advance\eqcount by 1
\ifnum\the\sectioncount=0
	\eqno({\oldstyle\the\eqcount})
\else
        \ifnum\the\appendixcount=0
                \eqno({\oldstyle\the\sectioncount}.{\oldstyle\the\eqcount})\fi
        \ifnum\the\appendixcount=1
                \eqno({\oldstyle A}.{\oldstyle\the\eqcount})\fi
        \ifnum\the\appendixcount=2
                \eqno({\oldstyle B}.{\oldstyle\the\eqcount})\fi
        \ifnum\the\appendixcount=3
                \eqno({\oldstyle C}.{\oldstyle\the\eqcount})\fi
\fi}
\def\multi{\global\advance\eqcount by 1}
\def\multieq#1#2{\xdef#1{{\old\the\eqcount#2}}
        \eqno{({\oldstyle\the\eqcount#2})}}
%%%%%%%%%%%%%%%%%%%%%%%%%%%%%%%%%%%%%%%%%%%%%%%%%%%%%%%%%%%%%%%%%
%                                                               %
%       Hyperrefs:                                        	%
%                                                               %
%%%%%%%%%%%%%%%%%%%%%%%%%%%%%%%%%%%%%%%%%%%%%%%%%%%%%%%%%%%%%%%%%
\newtoks\url
\def\Href#1#2{\catcode`\#=12\url={#1}\catcode`\#=\active#2}
\def\href#1#2{{#2}}
\def\hhref#1{{#1}}
%%%%%%%%%%%%%%%%%%%%%%%%%%%%%%%%%%%%%%%%%%%%%%%%%%%%%%%%%%%%%%%%%
%                                                               %
%       FORMAT:                                                 %
%                                                               %
%%%%%%%%%%%%%%%%%%%%%%%%%%%%%%%%%%%%%%%%%%%%%%%%%%%%%%%%%%%%%%%%%
\parskip=3.5pt plus .3pt minus .3pt
\baselineskip=14pt plus .1pt minus .05pt
\lineskip=.5pt plus .05pt minus .05pt
\lineskiplimit=.5pt
\abovedisplayskip=18pt plus 4pt minus 2pt
\belowdisplayskip=\abovedisplayskip
\hsize=14cm
\vsize=19cm
\hoffset=1.5cm
\voffset=1.8cm
\frenchspacing
\footline={}
\raggedbottom
%%%%%%%%%%%%%%%%%%%%%%%%%%%%%%%%%%%%%%%%%%%%%%%%%%%%%%%%%%%%%%%%%
%                                                               %
%       VARIOUS DEFINITIONS                                     %
%                                                               %
%%%%%%%%%%%%%%%%%%%%%%%%%%%%%%%%%%%%%%%%%%%%%%%%%%%%%%%%%%%%%%%%%

\def\ss{\scriptstyle}
\def\sss{\scriptscriptstyle}
\def\*{\partial}
\def\punkt{\,\,.}
\def\komma{\,\,,}

\def\={\!=\!}
\def\small#1{{\hbox{$#1$}}}
\def\half{\small{1\over2}}
\def\fraction#1{\small{1\over#1}}
\def\fr{\fraction}
\def\Fraction#1#2{\small{#1\over#2}}
\def\Fr{\Fraction}

\def\eg{{\tenit e.g.}}

\def\ie{{\tenit i.e.}}

\def\nl{\hfill\break\indent}
\def\nlni{\hfill\break}

\def\d{\delta}
\def\e{\varepsilon}

\def\Z{{\Bbb Z}}

\def\R{{\Bbb R}}

\def\Ham{{\cal H}}

\def\d{\partial}
\def\T{{\cal T}}

\def\s{\sigma}
\def\r{\varrho}

%%%%%%%%%%%%%%%%%%%%%%%%%%%%%%%%%%%%%%%%%%%%%%%%%%%%%%%%%%%%%%%%%%%%%%%%%%%%%

\ref\deWitHoppeNicolai{B. de Wit, J. Hoppe and H. Nicolai,
	{\xit ``On the quantum mechanics of supermembranes''},
	\NPB{305}{1988}{545}.}

\ref\deWitLuscherNicolai{B. de Wit, M L\"uscher and H. Nicolai,
	{\xit``The supermembrane is unstable''},
	\NPB{320}{1989}{135}.}

\ref\deWitMarquardNicolai{B. de Wit, U. Marquard and H. Nicolai,
	{\xit ``Area preserving diffeomorphisms and supermembrane 
	Lorentz invariance''}, \CMP{128}{1990}{39}.}

\ref\CederwallOpenMembrane{M. Cederwall, 
	{\xit ``Boundaries of {\xold11}-dimensional membranes''}, 
	\MPLA{12}{1997}{2641} [\hepth{9704161}].}

\ref\CederwallAffine{M. Cederwall, {\xit ``Open and winding membranes,
	affine matrix theory and matrix string theory''},
	\jhep{02}{12}{2002}{005} [\hepth{0210152}].}

\ref\CederwallLarsson{M. Cederwall and H. Larsson, {\xit ``M5-branes
	and matrix theory''}, in the proceedings of the International
	Seminar on Supersymmetries and Quantum Symmetries (SQS '03),
	Dubna, July 2003 and of the 9th Adriatic Meeting, Dubrovnik,
	Sept. 2003, \hepth{0312303}.}

\ref\HoravaWitten{P. Ho\v rava and E. Witten,
	{\xit ``Heterotic and type I string dynamics from eleven-dimensions''},
	\NPB{460}{1996}{506} [\hepth{9510209}];
	{\xit ``Eleven-dimensional supergravity on a manifold with boundary''},
	\NPB{475}{1996}{94} [\hepth{9603142}].}

\ref\FairlieZachos{D.B. Fairlie and C.K. Zachos, 
	{\xit ``Infinite dimensional algebras, sine brackets and 
	{\xit su}($\ss\infty$)''}, \PLB{224}{1989}{101}.}

\ref\KimRey{N. Kim and S.-J. Rey,
	{\xit ``M(atrix) theory on an orbifold and twisted membrane''},
	\NPB{504}{1997}{189} [\hepth{9701139}].}

\ref\EzawaMatsuoMurakami{K. Ezawa, Y. Matsuo and K. Murakami,
	{\xit ``Matrix regularization of open supermembrane: 
	towards M theory five-brane via open supermembrane''},
	\PRD{57}{1998}{5118} [\hepth{9707200}].}

\ref\BeckerBecker{K. Becker and M. Becker, {\xit ``Boundaries in M theory''},
	\NPB{472}{1996}{221} [\hepth{9602071}].}

\ref\Hoppe{J. Hoppe, {\xit ``Zero energy states in supersymmetric matrix models''},
	\CQG{17}{2000}{1101}, and references therein.}

\ref\Reviews{M.J. Duff, {\xit ``Supermembranes''}, \hepth{9611203};
	\nlni H. Nicolai and R. Helling, 
		{\xit ``Supermembranes and (M)atrix theory''}, \hepth{9809103};
	\nlni B. de Wit, {\xit ``Supermembranes and super matrix models''}, 
		\hepth{9902051}; 
	\nlni W. Taylor, {\xit ``(M)atrix theory: Matrix quantum mechanics 
		as fundamental theory''}, \nl\hepth{0101126};
	\nlni 
	T. Banks, {\xit ``TASI lectures on matrix theory''}, \hepth{9911068}.}

\ref\deWitPeetersPlefka{B. de Wit, K. Peeters and J. Plefka,
	{\xit ``Supermembranes with winding''}, 
	\PLB{409}{1997}{117} [\hepth{9705225}]}

\ref\BFSS{T. Banks, W. Fischler, S.H. Shenker and L. Susskind,
	{\xit ``M theory as a matrix model: a conjecture''}, 
	\PRD{55}{1997}{5112} [\hepth{9610043}].}

\ref\SethiStern{S. Sethi and M. Stern, {\xit ``D-brane bound states redux''},
		\CMP{194}{1998}{675} [\hepth{9705046}].}

\ref\KacSmilga{V.G. Kac and A.V. Smilga, 
		{\xit ``Normalized vacuum states in N=4 supersymmetric 
		Yang--Mills quantum mechanics with any gauge group''},
		\NPB{571}{2000}{515} [\hepth{9908096}].}

\ref\BST{E. Bergshoeff, E. Sezgin and P.K. Townsend,
	{\xit ``Supermembranes and eleven-dimensional supergravity''},
	\PLB{189}{1987}{75}; 
	{\xit ``Properties of the eleven-dimensional super membrane theory''},
	\AP{185}{1988}{330}.}

\ref\BSTT{E. Bergshoeff, E. Sezgin, Y. Tanii and P.K. Townsend,
	{\xit ``Super p-branes as gauge theories of 
	volume preserving diffeomorphisms''},
	\AP{199}{1990}{340}.}

\ref\BBSS{E. Bergshoeff, D.S. Berman, J.P. van der Schaar and
	P.Sundell, {\xit ``A non-commutative M-theory five-brane''},
	\NPB{590}{2000}{173} [\hepth{0005026}].}

\ref\BCGLNNS{D.S. Berman, M. Cederwall, U. Gran, H. Larsson,
	M. Nielsen, B.E.W. Nilsson and P. Sundell, {\xit
	``Deforma\-tion-independent open brane metrics and generalised
	theta parameters''}, \jhep{02}{02}{2002}{012} [\hepth{0109107}].}

\ref\ChuSezgin{C.-S. Chu and E. Sezgin, 
	{\xit ``M five-brane from the open supermembrane''},
	\JHEP{97}{12}{1997}{001} [\hepth{9710223}].}

\ref\FuchsSchweigert{J. Fuchs and C. Schweigert, 
	{\xit ``Symmetries, Lie algebras and representations''},
	Cambridge University Press, 1997.}

\ref\DVV{R. Dijkgraaf, E. Verlinde and H. Verlinde,
	{\xit ``Matrix string theory''}, \NPB{500}{1997}{43} [\hepth{9703030}].} 

\ref\SekinoYoneya{Y. Sekino and T. Yoneya, 
	{\xit ``From supermembrane to matrix string''},
	\NPB{619}{2001}{22} [\hepth{0108176}].}

\ref\Minic{Dj. Mini\'c, {\xit ``M-theory and deformation quantization''},
	\hepth{9909022}.}	

\ref\BraxMourad{Ph. Brax and J. Mourad, 
	{\xit ``Open supermembranes in eleven-dimensions''},
	\PLB{408}{1997}{142} [\hepth{9704165}];
	{\xit ``Open supermembranes coupled to M theory five-branes''},
	\PLB{416}{1998}{295} [\hepth{9707246}].}

\ref\Henningson{A. Gustavsson and M. Henningson, 
	{\xit ``A short representation of the six-dimensional (2,0) algebra''},
	\JHEP{01}{06}{2001}{054} [\hepth{0104172}].}

\ref\BermanEtAl{D.S. Berman, M. Cederwall, U. Gran, H. Larsson,
      M. Nielsen, B.E.W. Nilsson and P. Sundell
	{\xit ``Deformation independent open brane metrics and generalized
      theta parameters''},
	\jhep{02}{02}{2002}{012} [\hepth{0109107}].}

\ref\Schaar{J.P. van der Schaar, {\xit ``The reduced open membrane metric''},
	\jhep{01}{08}{2001}{048} [\hepth{0106046}].}

\ref\BermanPioline{D.S. Berman and B. Pioline, {\xit ``Open membranes,
ribbons and deformed Schild strings''}, \PRD{70}{2004}{045007} [\hepth{0404049}].}

\ref\Swain{J. Swain, {\xit ``On the limiting procedure by which
SDiff(T${}^{\sss2}$) and su($\ss\infty$) are related}, \hepth{0405002}.}

\ref\HoppeNambu{J. Hoppe, {\xit ``On M algebras, the quantisation of
Nambu mechanics, and volume preserving diffeomorphisms''},
\HPA{70}{1997}{302} [\hepth{9602020}].}

\ref\Awata{H. Awata, M. Li, Dj. Mini\'c and T. Yoneya, {\xit ``On the
quantization of Nambu brackets''}, \jhep{01}{02}{2001}{013}
[\hepth{9906248}].}

\ref\ZachosNambu{C. Zachos, {\xit ``Membranes and consistent
quantization of Nambu dynamics''}, \PLB{570}{2003}{82} [\hepth{0306222}].}

\ref\Zariski{G. Dito, M. Flato, D. Sternheimer and L. Takhtajan, {\xit
``Deformation quantization and Nambu mechanics''}, \CMP{183}{1997}{1} [\hepth{9602016}].}

\ref\Kulish{M. Chaichian, P.P. Kulish, K. Nishijima and A. Tureanu,
{\xit ``On a Lorentz-invariant interpretation of non-commutative
space-time and its implications on non-commutative QFT''},
\hepth{0408069}.}

\ref\Lawrence{R.J. Lawrence, {\xit ``On algebras and triangle
relations''}, in proceedings of ``Topological and geometrical methods
in field theory'', Turkku, May 1991.}

%%%%%%%%%%%%%%%%%%%%%%%%%%%%%%%%%%%%%%%%%%%%%%%%%%%%%%%%%%%%%%%%%%%%%%%%%%%%%

\evenheadtext={Martin Cederwall}
\oddheadtext={Thoughts on Membranes, Matrices and Non-Commutativity}

\line{
\epsfysize=1.7cm
\epsffile{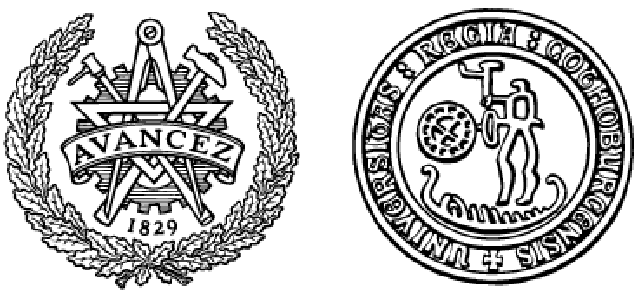}
\hfill}
\vskip-1.7cm
\line{\hfill G\"oteborg ITP preprint}
\line{\hfill hep-th/0410110}
\line{\hfill October, {\old2004}}
\line{\hrulefill}

\vfill

%\null\vskip5cm
\centerline{\sixteenhelvbold Thoughts on Membranes, Matrices} 
\vskip6pt
\centerline{\sixteenhelvbold and Non-Commutativity} 

\vskip8mm
\centerline{Invited talk presented at the conference ``Non-Commutative Geometry
and}
\centerline{Representation Theory in Mathematical Physics'', Karlstad, Sweden,
July 5-10, 2004}

\vskip16mm

\catcode`\@=11
\centerline{\twelvehelvbold Martin Cederwall%\foot\star
%{\eighttt martin.cederwall@fy.chalmers.se}
}
\catcode`\@=\active

\vskip8mm

\centerline{\it Department of Theoretical Physics}
%\vskip-1mm
\centerline{\it G\"oteborg University and Chalmers University of Technology }
%\vskip-1mm
\centerline{\it S-412 96 G\"oteborg, Sweden}

\vskip16mm

{\narrower\noindent%\baselineskip=.8\baselineskip\xrm 
\underbar{Abstract:} We review the passage from the supermembrane to 
matrix theory via a consistent truncation following 
a non-commutative deformation 
in light-cone gauge. Some indications are given that there should 
exist a generalisation of non-commutativity involving a three-index 
theta on the membrane, and we discuss some possible ways of 
investigating the corresponding algebraic structure.
\smallskip}

%\vskip5mm

\vfill

\vtop{\baselineskip=.8\baselineskip
\line{\hrulefill}
\catcode`\@=11
\line{\eighttt martin.cederwall@fy.chalmers.se\hfill}
\catcode`\@=\active
\line{\eighttt \hhref{http://fy.chalmers.se/tp/cederwall/}\hfill}
}

\eject

\section\Introduction{Introduction}String theory, or M-theory, which
seems a more proper name in the present context, possesses vacuum
solutions in which there are no perturbative excitations, and (thus)
no strings. Among these is the 11-dimensional Minkowski space (among
other solutions to 11-dimensional supergravity), in
which the effective low-energy excitations are described by
11-dimensional supergravity. In this background, M-theory allows the
existence of certain extended objects, namely the M2- and M5-branes,
which can be seen \eg\ by considering solutions to the supergravity
equations of motion. The M2-brane, or 11-dimensional supermembrane, is
electrically  charged with respect to the 3-form tensor field of the
supergravity  theory, and the M5-brane is magnetically charged (in
addition there are objects that carry gravitational charges, but these
will not be considered here). In the absence of strings, one may ask
if membranes may play the r\^ole of fundamental excitations of
M-theory, albeit non-perturbative. Such a speculation is supported by
the nature of the corresponding supergravity solution, which has a
singularity (the M5-brane is completely smooth, which is typical for a
truly solitonic object). It shows a difference, however, 
compared to fundamental string solutions in 10 dimensions
exhibiting naked singularities in the string frame, in that the singularity  is
surrounded by a horizon, probably indicating the non-perturbative
nature of membranes. There does not seem to exist a limit where
membrane states are perturbative.

Quantisation of membranes [\deWitHoppeNicolai] 
has been a challenging problem since their
theoretical discovery (for reviews of membranes and matrix theory, see
ref. [\Reviews]), and there is at this point no complete solution
of the problem. It is generally believed that a solution should
provide valuable insight into the non-perturbative nature of string
theory/M-theory, although presumably not in a background-independent
form. Note that since there is no perturbation theory, there is no
such distinction as between a first-quantised and a second-quantised
theory, and membrane quantisation has to be very different from string
quantisation (producing the one-string states used to build a
multi-string Fock space).

Most of the progress in this area of M-theory has been made in terms
of matrix theory. Matrix theory is in a well-defined way an
approximation, or even a consistent truncation, of membrane
theory. The first section of this talk will be 
devoted to a demonstration of the
relation between membranes and matrices. In this procedure,
non-commutativity plays a crucial (mathematical) r\^ole. 
In the subsequent sections,
we will discuss the physical relevance of this non-commutativity, and
also comment about its generalisations and other instances where
non-commutativity plays a r\^ole in M-theory.

\vfill\eject

\section\MembranesMatrices{From Membranes to Matrices}The action of an
M2-brane moving in a background which is a 
solution to the 11-dimensional supergravity
equations of motion is [\BST]
$$
S=-T\int_{{\cal M}_3}d^3\xi\sqrt{-g}+T\int_{{\cal M}_3}C\punkt\Eqn\Action
$$
Here, ${\cal M}_3$ is the membrane world-volume with coordinates 
$\{\xi^i,\,i=0,1,2\}$ and $T$ is the membrane tension, proportional to 
$\ell_P^{-3}$, $\ell_P$ being the 11-dimensional Planck length.
The fields $g$ and $C$ occurring in the action are pullbacks to the
membrane world-volume of the corresponding background superfields,
which means that the superspace coordinates $X^m$, $m=0,\ldots,11$ and
$\theta^\mu$, $\mu=1,\ldots,32$ are dynamical variables for the supermembrane.
In the following, we will, for sake of simplicity, 
exhibit only the bosonic variables, although
we will keep in mind that we are working with the supersymmetric
membrane, and comment at relevant places on implications of
supersymmetry. One crucial aspect of the supermembrane (as of any half-BPS
object) is
$\kappa$-symmetry, implying that the action (\Action) is invariant
under local translations in 16 of the 32 fermionic directions, so the
number of physical fermionic variables is 16. On-shell, we thus have 8
fermions and 8 bosons (the transverse fluctuations, the longitudinal
ones are killed by diffeomorphisms on ${\cal M}_3$).
We will also in flat 11-dimensional Minkowski space with vanishing
$C$-field. The $C$-field itself is relevant to another instance of
``non-commutativity''; this will be commented upon later.

In general, we should consider all possible topologies (\ie,
2-dimensional topologies at a given time) of the membrane, and keep in
mind the possibility of dynamical topology change. However, topology
seems to be quite unimportant in membrane theory. Here, we will only
give a heuristic argument for this. Consider a membrane consisting of
two surfaces connected by a very thin tube. The action for such a
configuration will in the limit where the tube becomes infinitesimally
thin only differ infinitesimally from the action for the configuration
where there are two membranes, meaning that (classical) membranes may
grow ``spikes'', connecting different parts or going towards infinity,
at no cost of energy. It also seems to indicate that a configuration
with one membrane already contains configurations with two or more
membranes.
As we will soon see, this property has to do with the existence of
flat directions in the potential. It is of course important to check
whether this property persists at the quantum level. It has been
proved that it does not for a bosonic membrane, due to zero-point
fluctuations in directions transverse to the flat directions, so that
quantum states of a bosonic membrane are localised. In a supersymmetric
situation however, the flat directions persist, due to cancellations
of zero-point energies between bosons and fermions. This gives an
indication that topology should be irrelevant for supermembranes, and
that ``first-quantised'' supermembrane Hilbert space is large enough
to contain an entire multi-particle Fock space. In this respect,
membranes behave very differently from strings. Strings have no flat
directions---a string stretched between two strings costs energy, and
topology change (change in the number of disconnected components)
represents interaction. The continuity of the supermembrane spectrum,
reflecting its multi-particle nature, was at first taken as a signal
that membrane quantisation produced undesired (for a first-quantised
theory) physical results [\deWitLuscherNicolai].
For these reasons, and for sake of simplicity, we will consider
membranes whose topology is a torus $T^2$.

The diffeomorphism constraints for a membrane are
$$
\eqalign{
\T_{i}\equiv P\cdot\d_iX\approx0\komma\cr
\Ham\equiv\fr{2T}P^2+\Fr T2\tilde g\approx0\komma\cr
}\eqn
$$
$\tilde g_{ij}=\d_iX^a\d_jX^b\eta_{ab}$ being the induced metric 
on the two-dimensional membrane space-sheet, 
and $\tilde g$ its determinant (from here on, $i,j,\ldots=1,2$ are
tangent indices on the space-sheet). It is conveniently rewritten in terms 
of the ``Poisson bracket'' on the space-sheet, 
$$
\{A,B\}=\e^{ij}\d_iA\d_jB\komma\Eqn\PoissonBracket
$$
as $\tilde g=\half\{X_a,X_b\}\{X^a,X^b\}$, so that the hamiltonian 
constraint takes the form
$$
\Ham=\fr{2T}P^2+\Fr T4\{X_a,X_b\}\{X^a,X^b\}\approx0\punkt\eqn
$$
We now want to go to a light-cone gauge. This is achieved by aligning
world-volume time with a light-like direction in target space, 
$X^+=x^++p^+\tau$, while the same component of the momentum is
demanded to be constant\foot\star{In this is many other equations,
one should insert an auxiliary density on the membrane space-sheet,
since the momentum, for example, is a density. On the torus, we can
choose it to be a constant.}, $P^+=p^+$. 
These gauge choices, together with the constraints, allow for solution
of $P^-$ and $X^-$, except for the constant mode of the latter (here,
and in the following, we set $T=1$):
$$
\eqalign{
P^-&={1\over2p^+}\left(P^IP^I+\half\{X^I,X^J\}\{X^I,X^J\}\right)\komma\cr
\d_iX^-&={1\over p^+}P^I\d_iX^I\komma\cr
}\Eqn\PmXm
$$
where the index $I$ now runs over the transverse light-cone directions
$I=1,\ldots,9$. $P^-$ is the light-cone hamiltonian. The physical
phase space is spanned by $X^I$, $P^I$, whose Dirac brackets, as usual
in light-cone gauge, are identical to the original 
Poisson brackets\foot\dagger{This is the ordinary dynamical Poisson bracket,
not to be confused with the structure introduces in
eq. (\PoissonBracket).},
$[X^I(\xi),P^J(\xi')]_{PB}=\delta^{IJ}\delta^{(2)}(\xi-\xi')$.
From the form of $\d_iX^-$ in eq. (\PmXm) it follows as an
integrability condition that $J\equiv\{X^I,P^I\}\approx0$. This is a
remaining gauge invariance, the area-preserving diffeomorphisms
(APD's) of the
membrane space-sheet [\BSTT,\deWitMarquardNicolai]. 

If we expand the fields in momentum eigenfunctions on $T^2$ labeled by
a pair of integers $k_i$, the Poisson bracket between two functions is
$$
\{e^{ik\cdot\xi},e^{ik'\cdot\xi}\}=-\epsilon^{ij}k_ik_je^{i(k+k')\cdot\xi}
\punkt\Eqn\FunctionPBAlgebra
$$
This is also the algebra (under $[\cdot,\cdot]_{PB}$) 
 between modes of the generator $J$ of APD's.
It is well-known that this algebra in a certain sense is isomorphic to 
``$u(\infty)$'' [\FairlieZachos]. 
Namely, a parametrisation of $u(N)$ generators in
terms of ``clock and shift matrices'' gives the $u(N)$ commutation
relations as 
$$
[T_k,T_{k'}]=-2i\sin({\pi\over N}\epsilon^{ij}k_ik_j)T_{k+k'}
\komma\Eqn\DeformedAlgebra
$$
for pairs of integers $k$, $k'$ ranging from 0 to $N$.
Clearly, eq. (\DeformedAlgebra) provides a ``regularisation'' of 
eq. (\FunctionPBAlgebra), in the sense
that, for large $N$, the modes with mode numbers $<<\sqrt{N}$ (low
enough to approximate the sine function with a linear function) behave the same
way (modulo rescaling) in both, while the number of generators in eq. 
(\DeformedAlgebra) is finite\foot\star{From the literature, one may
get the impression that there is some disagreement on the relation
between the algebra of APD's and $\ss u(\infty)$ (see {\xit e.g.}
ref. [\Swain]). 
This confusion can be attributed to
observations that if one keeps {\xit all} generators in
eq. (\DeformedAlgebra), {\xit i.e.}, also those beyond the scope of the linear
approximation, there is no way to obtain eq. (\FunctionPBAlgebra) as
an $\ss N\rightarrow\infty$ limit. In a physical interpretation, one has
to invoke a kind of low-energy approximation, the correct statement
being that, for any {\xit fixed} $\ss k$'s, the limit of the $\ss su(N)$
commutators is given by eq. (\FunctionPBAlgebra). 
Also, depending on topology, there may
be APD's not connected to the identity. With regard to the discussion
on topology, we believe it to be physically motivated not to consider
these.}.

It seems therefore physically motivated to regulate the algebra of
functions and of the APD's by replacing it by $u(N)$. Here we want to
advocate the view, however, that the best way of reaching that result
is by {\it consistent truncation}. A consistent truncation of a
dynamical system consists of a choice of a subset of modes, whose
excitations do not force the excitations of the remaining modes, \ie,
setting the remaining modes to zero is consistent with the equations
of motion of the system. This is clearly not achieved by approximating
the algebra as above. The procedure we advocate is instead, as
stressed \eg\ in ref. [\CederwallAffine], a
non-commutative deformation of the algebra of functions, 
followed by a consistent truncation. In that way, one will regain full
control over any algebraic property of the theory, like \eg\
supersymmetry.

The algebra of functions is deformed with a non-commutativity parameter
$\theta$ by $[\xi^1=\s,\xi^2=\r]=i\theta$ and Weyl ordering, so that
$$
f{\star}g=
fe^{{i\over2}\theta\e^{ij}\buildrel{\ss\leftarrow}\over\d\!\!{}_i
\buildrel{\ss\rightarrow}\over\d\!\!{}_j}g\punkt\eqn
$$
The Fourier modes of definite momentum then commute as
$$
[e^{ik_i\xi^i},e^{ik'_i\xi^i}]_{\star}=-2i\sin(\half\theta\epsilon^{ij}k_ik'_j)
e^{i(k_i+k'_i)\xi^i}
\komma\Eqn\StructureConstants
$$
If $\theta$ is ``rational'', $\theta=\Fr{2\pi q}N$ where $q$ and $N$ 
are coprime integers, the sine function
in the structure constants will have zeroes. The functions
$e^{isN\s}$ and $e^{irN\r}$ ($s,r\in\Z$) commute with all other 
functions, they are central elements in the algebra. This means that 
they can be consistently modded out from the algebra of function under 
the star product, since left and right multiplication coincide on all 
functions. If the star products by $e^{iN\s}$ and $e^{iN\r}$ are 
identified with the identity operator, one obtains the equivalences
$$
\eqalign{
e^{i(k+N)\s+il\r}&\approx (-1)^le^{ik\s+il\r}\komma\cr
e^{ik\s+i(l+N)\r}&\approx (-1)^ke^{ik\s+il\r}\punkt\cr
}\eqn
$$
The star-commutator algebra after this ``consistent truncation'' is 
identical to eq. (\DeformedAlgebra), $u(N)$.

To summarise, matrix theory is obtained from membrane theory in two
steps, the first one being a non-commutative deformation with 
$\{\cdot,\cdot\}\rightarrow[\cdot,\cdot]_{\star}$, the second one a
consistent truncation with 
$[\cdot,\cdot]_{\star}\rightarrow[\cdot,\cdot]_{N\times N}$.
A less strict and more intuitive way of explaining the need for a
deformation in order to enable a consistent truncation would be that
the (rational) 
deformation effectively introduces a discretisation of the torus,
in which the APD's act by finite translations, while they in the
undeformed case act by infinitesimal ones. In this way the deformed
algebra will contain elements that translate all the way around a
periodic directions, while this does not happen in the undeformed
case, accounting for the impossibility of
finding an $su(N)$ subalgebra of ``$su(\infty)$''.

Modulo constants, the light-cone hamiltonian for the matrix theory is
$$
P^-\sim{1\over2p^+}\hbox{Tr}\left(P^IP^I+\half[X^I,X^J][X^I,X^J]\right)
+\hbox{terms with fermions}\komma\Eqn\MHam
$$
where the $X$'s and $P$'s are $N\times N$ matrices,
which is proportional to the hamiltonian for the dimensional reduction
to $D=1$ of $D=10$ SYM with gauge group U($N$).

The flat directions of the potential in eq. (\MHam) form the
``classical moduli space''. This is easily seen by diagonalisation to be 
$\R^9\times(\R^{9(N-1)}/P_N)$, $P_N$ being the permutation group of
$N$ objects. So, while the diagonal matrices can be seen as
parametrising the motion of $N$ ``partons'', the off-diagonal entries
represent interactions. It should be stressed that this multi-particle
interpretation of the one-membrane Hilbert space provides an
interesting alternative, in terms of ordinary quantum mechanics, to
quantum field theory. For the supersymmetric membrane, as mentioned
earlier, the existence of the flat directions persists at the quantum
level. The full non-linear problem seems very difficult, so far only
the form of the asymptotic scattering states (far out in the valleys)
has been found exactly, along with low-order series expansions in
velocities for interactions [\Hoppe]. Still, index theorems ensure that a
unique ground state exists [\SethiStern,\KacSmilga]. 
This is crucial for the connection to
M-theory; then the fermionic zero-modes generate the $2^8$ physical
fields of 11-dimensional supergravity\foot\dagger{The 16 fermionic
zero-modes act as gamma matrices of $\ss so(16)$. Under the ``Hopf
reduction'' $\ss so(16)\rightarrow so(9)$, the fermionic Hilbert
space, an $\ss so(16)$ spinor, decomposes as $\ss
128_s\rightarrow44\oplus84$ (the metric and $\ss C$-field)
and $\ss 128_c\rightarrow 128$ (the gravitino).}.

The corresponding procedure with deformation and consistent truncation
is applicable in other situations, \eg\ when the membrane winds a
compactified circle, in which case matrix string theory is produced 
[\SekinoYoneya,\CederwallAffine],
or in backgrounds with $C$-field, which gives rise to a deformation of
the SYM theory [\CederwallLarsson]. 

\section\CField{The {\twelveit C}\hskip2pt-Field 
and Non-Commutativity}In string theory,
field theories on D-branes, in the presence of a background 2-form
$B$-field, can be seen as non-commutative. The non-commutativity
parameter is directly related to the $B$-field. One way of deriving this
is to consider the mixed Dirichlet/Neumann boundary condition for a
string ending on the D-brane as a set of constraints and deriving the
corresponding Dirac brackets for the string end-point coordinates. In
11 dimensions, membranes may end on M5-branes in an analogous way, and
one would expect a similar mechanism to be at work. The situation is
complicated by the limited knowledge of the theory on M5-branes
relevant for the situation. A simplistic approach using the membrane
action yields a non-commutative loop space with a three-index
theta [\BBSS]. The precise form of the non-commutativity is not known, due to
the non-linearity of membrane dynamics. Similar results, with
three-index thetas, have been obtained from supergravity
solutions [\BCGLNNS,\Schaar,\BermanPioline]. The precise meaning is so far 
much more elusive than in
string theory.

\section\TwoThree{From Two-Index to Three-Index?}Both instances of
non-commutativity in membrane theory, and their relation to string
theory, can be taken as indicating that an algebraic structure should
be lifted/generalised to some structure involving three indices. In
the case of the previous section, the string theory $B$-field comes
from a $C$-field in 11 dimensions. In the case of non-commutativity on
the membrane, as described in section (\MembranesMatrices), $\theta$
is introduces as a mathematical tool, and unless it  has a more
profound physical meaning (see the following section), one might
regards it as specific to light-cone gauge. It is nevertheless
tempting to speculate in ``covariant'' three-index structures
involving the entire world-volume, not only the space-sheet. 

Some authors have argued that a formulation in terms of Nambu brackets 
should be relevant for membrane dynamics
[\HoppeNambu,\Awata,\ZachosNambu]. 
The Nambu bracket is the
straight-forward generalisation of the Poisson bracket to a
``three-product'',
$\{A,B,C\}=\epsilon^{ijk}\d_iA\d_jB\d_kC$. In addition to being
completely antisymmetric, it obeys a derivation property (like the
Poisson bracket) and a
``Fundamental Identity'' (FI, generalising the Jacobi identity for Poisson
brackets):
$$
\eqalign{
\{AB,C,D\}&=A\{B,C,D\}+\{A,C,D\}B\komma\cr
\{A,B,\{C,D,E\}\}&=\{\{A,B,C\},D,E\}+\{C,\{A,B,D\},E\}\cr
&+\{C,D,\{A,B,E\}\}
\punkt\cr
}\eqn
$$
If one wishes to deform the Nambu bracket, abstract algebra may not
give enough information on how to do this. One should think about the
physical meaning of \eg\ the FI, in order to know whether it should
hold in the deformed case. Of course, if the Nambu bracket $\{A,B,X\}$
is interpreted as a transformation with parameters $A$ and $B$ on $X$
(three-dimensional volume-preserving diffeomorphisms),
the FI is necessary for the closure of the transformations to a Lie algebra.
On the other hand, although the membrane action may be rewritten in
terms of Nambu brackets, the symmetries of the membrane are full
diffeomorphisms and not expressible in terms of Nambu brackets, which
makes the situation somewhat unclear.
An attempt to deform the Nambu bracket with a three-index $\theta$
fails, if one tries a na\"\i ve expansion in momenta (the same kind of
deformation that deforms the Poisson bracket to a star
commutator). Other methods have been devised, see \eg\ ref. [\Zariski], which
relies on factorisation of polynomials.

Our opinion is that in order to get further on this track, one would
need some intuition concerning the physical meaning of a
three-structure. A relaxation of some of the the identities
[\Awata] 
would need to be
motivated and replaced by some weaker identity, in order not to end up
in a situation where, for example, it is no longer possible to realise
``symmetries'' infinitesimally as operators on a Hilbert space (unless
one has clear reasons for this).

An inventory of other mathematical ``three-structures'' include \eg\
Jordan triple products (with a defining identity which, in spite of different
symmetry properties, can be written in a form identical to the FI of
the Nambu bracket) and the 3-algebras of ref. [\Lawrence]. It is not clear why
either of these structures should be relevant to membranes, but at a
purely mathematical level there are connections between these and the
brackets of ref. [\Awata].

\section\Lorentz{Symmetries of Non-Commutative Membranes?}The 
membrane ``partons'' lack some properties to make them independent
particles in 11 dimension. Even if they move independently in the 9
transverse directions, there is only one collective momentum $p^+$. On
reduction to $D=10$ matrix theory [\BFSS], 
and type IIA, so called ``discrete light-cone
quantisation'' (DLCQ, which will not be discussed in detail here, see \eg\
ref. [\Reviews]), they become D0-branes. In that process the size $N$ of the
matrices is proportional to $p^+$. 
We now come to some serious questions concerning the physical
interpretation of the truncated theory in 11 dimensions. Even if we
have kept as much as possible of the algebraic structure of the
theory, so that we are ensured that any linearly realised symmetry of
the original theory will pertain, some symmetries are non-linearly
realised in light-cone gauge. This happens especially for for the
dynamical generators $p^-$ and $j^{-I}$ of the Poincar\'e algebra and half of
the supersymmetry generators. 

Several authors have investigated the Poincar\'e invariance of the
truncated supermembrane, with the conclusion that it is broken by the
truncation, but recovered in the $N\rightarrow\infty$ limit. This
should not be surprising: it does not seem reasonable to expect such a
symmetry in an interacting theory when particle number is restricted
(remember that the dynamical generators are exactly the ones that
transform out of the initial value surface). The identification of
$p^+$ with particle number in the DLCQ procedure also indicates this. 
The relation between $N$ and the non-commutativity parameter $\theta$
then seems to indicate that if there is a covariant version of matrix
theory, one should allow $\theta$ to transform. This would be possible
if it happens after the deformation, but before the truncation. We
think that another advantage of the two-step consistent truncation
procedure may be that it allows this kind of line of thought.

One could discuss at length whether such an attempt would be
successful. For example, $\theta$ is a scalar from
target space point of view, so its introduction might still be
consistent with Poincar\'e symmetry. Also, from a world-volume
perspective, introduction of non-commutativity does not break
rotational symmetry, but deforms it (see the talk by P. Kulish at this
conference [\Kulish]). 
Probably, only a direct construction of the dynamical generators,
or a proof of the impossibility thereof will settle the question. Work
in this direction is under way. Here we would only like to mention
that the replacement of APD's by gauge transformations may give a
clue. Namely, if we still expect the remaining gauge symmetry to arise
as an integrability condition on $X^-$, we are lead to another
expression for $X^-$ involving a superposition of different values of
$\theta$, 
$$
\d_iX^-={1\over2p^+\theta}\int_0^\theta d\theta'
(\d_iX^I{\star}_{\theta'}P^I+P^I{\star}_{\theta'}\d_iX^I)\punkt\eqn
$$
which is easily verified by noting that differentiating with respect
to $\theta$ brings down two derivatives. Whether insertion of this new
relation, together with other suitable modifications, 
may lead to Lorentz covariance remains to be seen.

\acknowledgements
The author wants to thank the organisers of the conference for a
stimulating experience and Jens Hoppe and Daniel Sternheimer for
comments and information concerning their work.

\refout
\end